# New Silicon Carbide (SiC) Hetero-Junction Darlington Transistor

M. Jagadesh Kumar, *Senior Member, IEEE*, and Amit Sharma

*Abstract*--Basic SiC bipolar transistors have been studied in the past for their applications where high power or high temperature operation is required. However since the current gain in SiC bipolar transistors is very low and therefore, a large base drive is required in high current applications. Therefore, it is important to enhance the current gain of SiC bipolar transistors. Using two dimensional mixed mode device and circuit simulation, for the first time, we report a new Darlington transistor formed using two polytypes 3C-SiC and 4H-SiC having a very high current gain as a result of the heterojunction formation between the emitter and the base of transistor. The reasons for the improved performance are analyzed.

## I. INTRODUCTION

SiC has been a material of choice for high temperature and high power applications due to its superior material as well as electrical properties over Si and GaAs and the availability of different polytypes in SiC. Being a wide band gap material it is suitable for high temperature applications where Si does not maintain its semiconductor behavior due to an increase in the minority carrier concentration at elevated temperatures. 4H-SiC BJTs have been extensively studied in the recent years however their main drawback is the low current gain [1, 2]. Darlington transistor is an attractive way to enhance the current gain of the transistor and there are reports regarding the monolithic [3,4] and hybrid [5] SiC Darlington transistors. The maximum current gain achieved in the case of monolithic Darlington transistor is ~ 2000 [3] whereas in hybrid case the large signal current gain is ~ 430 [5]. In this paper we report a Darlington transistor with a wide bandgap emitter having a very high current gain of the order of ~ $10^4$ obtained by implementing the concept of the heterojunction to increase the current gain of the Darlington transistor using 4H-SiC with a bandgap of 3.2 eV for the emitter and 3C-SiC with a bandgap of 2.2 eV for the base and the collector regions. We have used two dimensional mixed mode device and circuit simulation [6] to design and analyze the proposed structure.

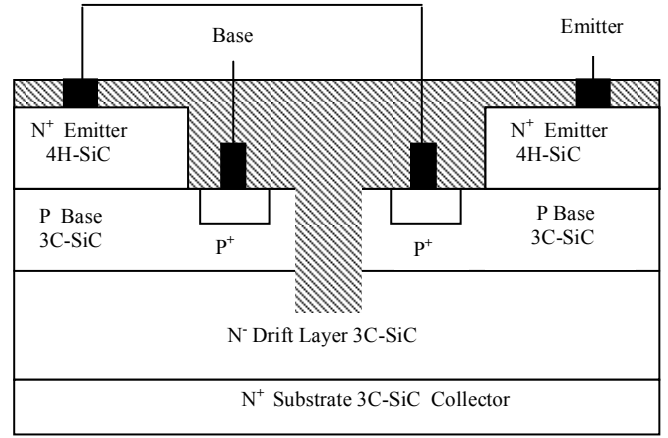

Fig. 1. Schematic corss-section of the hetero-junction Darlington transistor.

## II. DEVICE STRUCTURE AND PARAMETERS

The schematic cross-section of the proposed structure is shown in the Fig. 1. The 4H-SiC emitter shown in Fig. 1 can be epitaxially grown over 3C-SiC which results in the formation of the heterojunction between the emitter and the base region. The thickness and doping of the base and N-drift layer are 0.5 µm, $1 \times 10^{18}$ cm$^{-3}$ and 20 µm, $1 \times 10^{17}$ cm$^{-3}$ respectively. The emitter region thickness is 0.5 µm and the doping is $1 \times 10^{19}$ cm$^{-3}$. The emitter trench required for the effective isolation of the two BJTs as shown in Fig.1 and can be formed using the reactive ion etching as suggested in the literature [4]. The material parameters used in simulation have been taken from the literature [7, 8]. The extrinsic base P$^+$ regions have a doping of $10^{19}$ cm$^{-3}$ whereas the collector contact is taken from the bottom. Emitter of the first stage BJT is shorted to the base of the second stage BJT using a metal line over the oxide.

## III. RESULTS AND DISCUSSION

We have simulated the structure shown in Fig. 1 above using the two dimensional mixed mode device and circuit simulator [6]. The Gummel plot for the SiC Darlington transistor without heterojunction is compared with that of the proposed structure with heterojunction as shown in Fig. 2. We notice that due to the heterojunction at the emitter-base junction, the collector current of the proposed Darlington transistor is significantly larger than the collector current of the conventional SiC Darlington without the heterojunction.

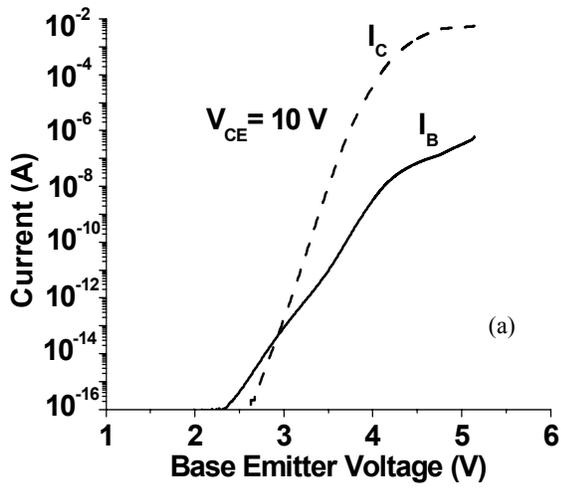

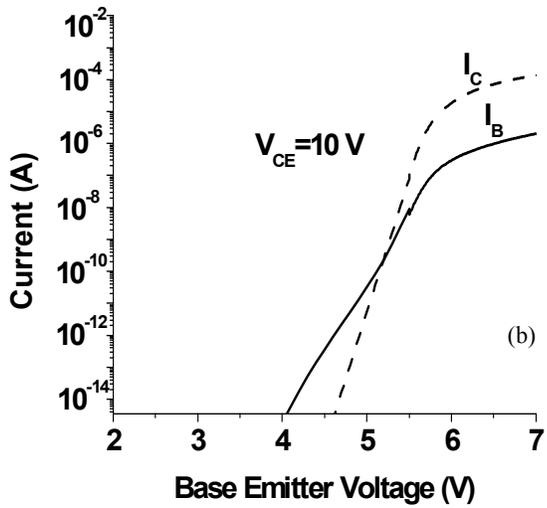

Fig. 2. Gummel plots for the SiC Darlington transistor (a) without heterojunction and (b) with heterojunction.

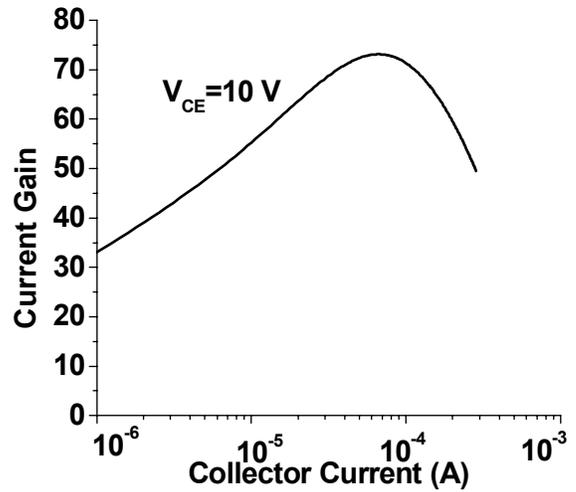

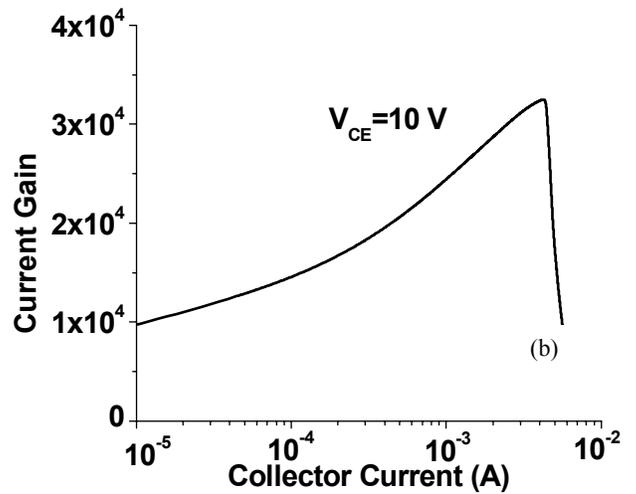

Fig. 3. Current gain versus collector current for the SiC Darlington transistor (a) without heterojunction and (b) with heterojunction.

This is expected to result in a significantly enhanced current gain over a wide range of collector current.

The common emitter current gain versus collector current for the SiC Darlington transistor (a) without heterojunction is compared with that of (b) the proposed structure with heterojunction as shown in Fig. 3. The maximum current gain for the proposed structure turns out to be of the order of $10^4$ which is very high as compared to the earlier reported current gains in the literature.

The variation of current gain with temperature for the proposed structure is studied to ensure the thermal stability of the device at elevated temperatures and is shown in Fig. 4. We notice that when the ambient temperature increases from 300 K to 400 K, there is only a marginal increase (10 -15 %) in current gain making this device suitable for use in high temperature applications. Most silicon transistors will fail at elevated temperatures due to a significant increase in current gain and the resultant thermal runaway.

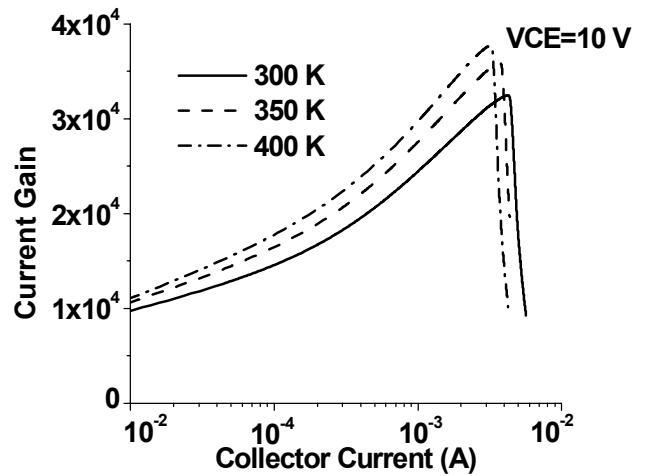

Fig. 4. Effect of temperature on the current gain of the heterojunction Darlington transistor.

## IV. CONCLUSION

SiC is a wide bandgap semiconductor having different polytypes with superior electrical properties as compared to Si. Availability of different polytypes in SiC makes it an excellent material for the heterostructure devices such as the one reported in this paper. For the first time the concept of heterojunction, in this case between two different SiC polytypes, has been applied to the power devices. Our results demonstrate that a significantly enhanced current gain can be achieved for a heterojunction Darlington transistor leading to a significant reduction in the base drive current in switching circuits. The proposed Darlington transistor will be a suitable candidate for high temperature applications since it is free of the gate oxide problems encountered in SiC MOSFETs and is thermally stable at higher temperatures.

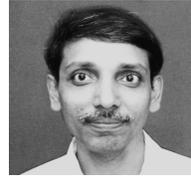

**M. Jagadesh Kumar** (SM'1999) was born in Mamidala, Nalgonda District, Andhra Pradesh, India. He received the M.S. and Ph.D. degrees in electrical engineering from the Indian Institute of Technology, Madras, India. From 1991 to 1994, he performed post-doctoral research in modeling and processing of high-speed bipolar transistors with the Department of Electrical and Computer Engineering, University of Waterloo, Waterloo, ON, Canada. While with the University of Waterloo, he also did research on amorphous silicon TFTs. From July 1994 to December 1995, he was initially with the Department of Electronics and Electrical Communication Engineering, Indian Institute of Technology, Kharagpur, India, and then joined the Department of Electrical Engineering, Indian Institute of Technology, Delhi, India, where he became an Associate Professor in July 1997 and a Full Professor in January 2005. His research interests are in Silicon Nanoelectronics, VLSI device modeling and simulation, integrated-circuit technology, and power semiconductor devices. He has published extensively in the above areas with more than 110 publications in refereed journals and conferences. His teaching has often been rated as outstanding by the Faculty Appraisal Committee, IIT Delhi.

Dr. Kumar is a *Fellow* of Institution of Electronics and Telecommunication Engineers (IETE), India and a *Senior Member* of IEEE. He is on the editorial board of *Journal of Nanoscience and Nanotechnology* and also on the Editorial Board of *IETE Journal of Research* as a subject area Honorary Editor for Electronic Devices and Components. He has reviewed extensively for different journals including *IEEE Trans. on Electron Devices, IEEE Trans. on Device and Materials Reliability and IEEE Electron Device Letters*. He was Chairman, Fellowship Committee, *The Sixteenth International Conference on VLSI Design,* January 4-8, 2003, New Delhi, India. He was Chairman of the Technical Committee for High Frequency Devices, *International Workshop on the Physics of Semiconductor Devices,* December 13-17, 2005, New Delhi, India.

**Amit Sharma** has completed his M.Tech degree in Solid State Materials in the Department of Physics, Indian Institute of Technology, Delhi. His research interests are design and simulation of Silicon Carbide power devices.